\documentclass[a4paper]{article}
\usepackage{INTERSPEECH2019}
\usepackage{xcolor}
\usepackage{float}
\usepackage[colorlinks=true,linkcolor=black,anchorcolor=black,citecolor=black,filecolor=black,menucolor=black,runcolor=black,urlcolor=black]{hyperref}

\title{BUT System Description to VoxCeleb Speaker Recognition Challenge 2019}

\name{
Hossein Zeinali,
Shuai Wang,
Anna Silnova,
Pavel Mat\v{e}jka,
and Old\v{r}ich Plchot
}
\address{Brno University of Technology, Speech@FIT and IT4I Center of Excellence, Brno, Czechia}

\email{\{zeinali,isilnova,matejkap,iplchot\}@fit.vutbr.cz, wsstriving@gmail.com}

\begin{document}

\maketitle

\begin{abstract}
In this report, we describe the submission of Brno University of Technology (BUT) team to the VoxCeleb Speaker Recognition Challenge (VoxSRC) 2019. We also provide a brief analysis of different systems on VoxCeleb-1 test sets. Submitted systems for both Fixed and Open conditions are a fusion of 4 Convolutional Neural Network (CNN) topologies. The first and second networks have ResNet34 topology and use two-dimensional CNNs. The last two networks are one-dimensional CNN and are based on the x-vector extraction topology. Some of the networks are fine-tuned using additive margin angular softmax.  Kaldi FBanks and Kaldi PLPs were used as features. The difference between Fixed and Open systems lies in the used training data and fusion strategy. The best systems for Fixed and Open conditions achieved 1.42\:\% and 1.26\:\% ERR on the challenge evaluation set respectively.
\end{abstract}


\section{Introduction}
\label{sec:intro}

As mentioned in the abstract, this document describes the Brno University of Technology (BUT) team submissions for the VoxCeleb Speaker Recognition Challenge (VoxSRC) 2019. This was the first challenge using VoxCeleb dataset. The challenge has two separate tracks: Fixed and Open. In the Fixed condition, participants can only use the development part of VoxCeleb-2 as training data while in the Open condition, they can use any data that they want.

Based on the success of Deep Neural Network (DNN) based embedding in speaker recognition, all of our systems are DNN based. One-dimensional Convolutional Neural Network (CNN) in a well-known x-vector extraction  topology~\cite{snyder2018x} was our first system for this task. We did several changes in the x-vector topology such as using more neurons and adding residual connections for enhancing its performance. 

The second approach to DNN based embedding extraction uses well known ResNet34 topology where several 2-dimensional CNNs are used in a very deep structure. Using residual connections in the ResNet helps the robustness of its training;  this network has achieved very good performance in various tasks. 

The rest of this document is organized as follows: in Section~\ref{sec:setup}, we first describe the setup for the challenge. In Section~\ref{sec:xvector}, the systems based on x-vector and ResNet34 DNN will be explained. Backends and fusion strategies are outlined in Section~\ref{sec:backend} and finally the results and analysis are presented in Section~\ref{sec:results}.

\section{Experimental Setup}
\label{sec:setup}

\subsection{Training data, Augmentations}
\label{HXV:data}

For all fixed systems, we used development part of VOXCELEB-2 dataset~\cite{Nagrani18VoxC} for training. This set has 5994 speakers spread over 145 thousand sessions (distributed in approx. 1.2 million speech segments). For training DNN based embeddings, we used original speech segments together with their augmentations. The augmentation process was based on the Kaldi recipe\footnote{\url{https://github.com/kaldi-asr/kaldi/tree/master/egs/sre16/v2}} and it resulted in  additional 5 million segments belonging to the following categories:

\begin{itemize}
 \setlength\itemsep{-0.2em}
    \item Reverberated using RIRs\footnote{\url{http://www.openslr.org/resources/28/rirs_noises.zip}}
    \item Augmented with Musan\footnote{\url{http://www.openslr.org/17/}} noise 
    \item Augmented with Musan music 
    \item Augmented with Musan babel
\end{itemize}

For Open condition, we tried to add more data to the VoxCeleb-2 development set. We first added the development part of VoxCeleb-1 with around 1152 speakers. The PLP-based systems were trained using this setup (i.e. VoxCeleb 1+2). For other open systems, we also used 2338 speakers from LibriSpeech dataset~\cite{LibriSpeech} and 1735 speakers from DeepMine dataset~\cite{zeinali2018deepmine}. For all training data, we first discarded utterances with less than 400 frames (measured after applying the VAD). After that, all speakers with less than 8 utterances (including  augmentation data) were removed.

\subsection{Development datasets}

We use the development data provided by the organizers\footnote{\url{http://www.robots.ox.ac.uk/~vgg/data/voxceleb/vox2.html}}.  Instead of using all the 6 trial lists, we only report our results on the cleaned versions: cleaned VoxCeleb1, cleaned Voxceleb1-E (extended) and cleaned Voxceleb1-H (hard). VoxCeleb1 test set is denoted as Voxceleb1-O (O for ``original'').

\subsection{Input features}
\label{HXV:vad_features}

We use different features for several systems with the  following settings:

\begin{itemize}
 \setlength\itemsep{-0.2em}
    \item {\bf 30-dimensional Kaldi PLP} - 16kHz, frequency limits 20-7600Hz, 25ms frame length, 40 filter-bank channels, 30 coefficients
    \item {\bf 40-dimensional Kaldi FBank} - 16kHz, frequency limits 20-7600Hz, 25ms frame length, 40 filter-bank channels
\end{itemize}

Kaldi PLPs and FBanks are subjected to short time mean normalization with a sliding window of 3 seconds.

\begin{table*}[tb]
\renewcommand{\arraystretch}{0.85}
\centering
\caption{\label{tab:xv_topology}x-vector topology proposed in \cite{snyder2019speaker}. $K$ in the first layer indicates different feature dimensionalities, $T$ is the number of training segment frames and $N$ in the last row is the number of speakers.}
\vspace{-2mm}
\begin{tabular}{c|c|c||c|c}\toprule
 \textbf{Layer} &  \multicolumn{2}{c||}{\textbf{Standard DNN}} &  \multicolumn{2}{c}{\textbf{BIG DNN}} \\
 & \textbf{Layer context} & \textbf{(Input) $\times$ output} & \textbf{Layer context} & \textbf{(Input) $\times$ output}\\ \cmidrule{1-5}
frame1 & $[t-2, t-1, t, t+1, t+2]$    & (5 $\times K$) $\times$ 512 & $[t-2, t-1, t, t+1, t+2]$    & (5 $\times K$) $\times$ 1024 \\
frame2 & $[t]$                        & 512 $\times$ 512 & $[t]$    & 1024 $\times$ 1024\\
frame3 & $[t-2, t, t+2]$              & (3 $\times$ 512) $\times$ 512 & $[t-4, t-2, t, t+2, t+4]$    & (5 $\times$ 1024) $\times$ 1024\\
frame4 & $[t]$                        & 512 $\times$ 512 &$[t]$                        & 1024 $\times$ 1024 \\
frame5 & $[t-3, t, t+3]$              & (3 $\times$ 512) $\times$ 512 & $[t-3, t, t+3]$              & (3 $\times$ 1024) $\times$ 1024\\
frame6 & $[t]$                        & 512 $\times$ 512 & $[t]$                        & 1024 $\times$ 1024  \\
frame7 & $[t-4, t, t+4]$              & (3 $\times$ 512) $\times$ 512 & $[t-4, t, t+4]$              & (3 $\times$ 1024) $\times$ 1024\\
frame8 & $[t]$                        & 512 $\times$ 512 & $[t]$                        & 1024 $\times$ 1024  \\
frame9 & $[t]$                        & 512 $\times$ 1500 & $[t]$                        & 1024 $\times$ 2000 \\
stats pooling & $[0, T]$              & 1500 $\times$ 3000  & $[0, T]$ & 2000 $\times$ 4000 \\
segment1 & $[0, T]$                      & 3000 $\times$ 512 & $[0, T]$ & 4000 $\times$ 512\\
segment2 & $[0, T]$                      & 512 $\times$ 512 & $[0, T]$ & 512 $\times$ 512\\
softmax  & $[0, T]$                      & 512 $\times N$ & $[0, T]$ & 512 $\times N$ \\
\bottomrule
\end{tabular}
\vspace{-3mm}
\end{table*}

\section{DNN based Systems}
\label{sec:xvector}

All Deep Neural Network (DNN) based embeddings used Energy-based VAD from Kaldi SRE16 recipe\footnote{We did not find a significant impact on performance when using different VAD within the DNN embedding paradigm and it seems that a simple VAD from Kaldi performs very well for DNN embedding in various channel conditions.}. For this challenge, we use two different embeddings: 

\subsection{x-vectors}
The first one is the well-known TDNN based x-vector topology. All its variants were  trained with Kaldi toolkit~\cite{povey2011kaldi} using SRE16 recipe with the following modifications: 

\begin{itemize}
 \setlength\itemsep{-0.2em}
    \item Using different feature sets (PLP, FBANK)
    \item Training networks with 6~epochs (instead of 3). We did not see any considerable difference with more epochs.
    \item Using modified example generation - we used 200~frames in all training segments instead of randomizing it between 200-400 frames. We have also  changed the training examples generation so that it is not random and uses almost all available speech from all training speakers.
    \item We used a bigger network~\cite{villalba2018jhu_mit} with more neurons in TDNN layers. Table~\ref{tab:xv_topology} shows a detailed description of the network.
    \item The BIG-DNN in Table~\ref{tab:xv_topology} was used for two PLP-based systems (i.e. systems 4 and 7 in Table~\ref{tab:results}). For other TDNN-based networks, we found that adding residual connections to the frame-level part of the network improves their performance. Therefore, other TDNN based networks used residual connections.
\end{itemize}

\subsection{ResNet34}
The second DNN embedding is based on the well-known ResNet34 topology~\cite{he2016deep}. This network uses 2-dimensional features as input and processes them using 2-dimensional CNN layers. Inspired by x-vector topology,  both mean and standard deviation are used as statistics. The detailed topology of the used ResNet is shown in Table~\ref{tab:resnet34}. We named the embedding extracted from ResNet as {\it ``r-vector''}. All ResNet networks were trained using SGD optimizer for 3 epochs using PyTorch. Similarly as in our previouos work  in TensorFlow~\cite{zeinali2019improve}, we found that L2-Regularization is useful here too.

\begin{table}[t]
  \renewcommand{\arraystretch}{1}
    \centering
    \caption{\label{tab:resnet34} The proposed ResNet34 architecture. $N$ in the last row is the number of speakers. The first dimension of the input shows number of filter-banks and the second dimension indicates the number of frames.}
    \vspace{-2mm}
    \setlength\tabcolsep{2pt}
    \begin{tabular}{l c c}
        \toprule
        \toprule
        \textbf{Layer name}   & \textbf{Structure}          & \textbf{Output} \\
        \midrule
        Input                 & --                          & 40 $\times$ 200 $\times$ 1  \\
        Conv2D-1              & 3 $\times$ 3, Stride 1      & 40 $\times$ 200 $\times$ 32 \\
        \midrule
        ResNetBlock-1         & $\begin{bmatrix} 3 \times 3, 32  \\ 3 \times 3, 32  \end{bmatrix} \times 3$  , Stride 1& $40\times 200 \times32$  \\
        ResNetBlock-2         & $\begin{bmatrix} 3 \times 3, 64  \\ 3 \times 3, 64  \end{bmatrix} \times 4$, Stride $2$ & $20 \times 100 \times 64$  \\
        ResNetBlock-3         & $\begin{bmatrix} 3 \times 3, 128 \\ 3 \times 3, 128 \end{bmatrix} \times 6$, Stride $2$ & $10 \times 50  \times 128$ \\
        ResNetBlock-4         & $\begin{bmatrix} 3 \times 3, 256 \\ 3 \times 3, 256 \end{bmatrix} \times 3$, Stride $2$ & $5  \times 25  \times 256$ \\
        \midrule
        StatsPooling          & --                & $10 \times 256$                 \\
        Flatten               & --                & $2560$                            \\
        \midrule
        Dense1                & --                & $256$                            \\
        Dense2 (Softmax)      & --                & $N$                              \\
        \midrule
        Total                 & --                & --                             \\
        \bottomrule
        \bottomrule
    \end{tabular}
    \vspace{-3mm}
\end{table}

\subsection{Fine-tuning networks with additive angular margin loss}

Additive angular margin loss (denoted as `AAM loss') was proposed for face recognition~\cite{deng2019arcface} and introduced to speaker verification in \cite{xiang2019margin}. Instead of training the AAM loss from scratch, we directly fine-tune a well-trained NN supervised by normal Softmax. To be more specific, all the layers after the embedding layer are removed (for both the ResNet and TDNN structure), then the remaining network is be fine-tuned using the AAM loss. For more details about AAM loss, see \cite{deng2019arcface} and~\cite{xiang2019margin}, $s$ is set to $30$ and $m$ is set to $0.2$ in all the experiments.

\section{Backend}
\label{sec:backend}

\subsection{Gaussian PLDA}

We used 500k randomly-selected utterances from VoxCeleb 2 for training the PLDA backend. We train it on embeddings extracted from the original utterances only, no augmented data was  used for training the backend. X-vectors were centered using the training data mean. Then, we applied LDA not reducing the dimensionality of the data. 
Finally, we did length normalization. Speaker and channel subspace size was set to 312.

\subsection{Cosine distance}

For ResNet embedding extractor ({\it r-vectors}) fine-tuned with additive angular margin loss, we performed simple cosine distance scoring. There was no preprocessing of the 256 or 160-dimensional embeddings except for centering. The centering mean was computed on 500k original VoxCeleb 2 utterances (the same data we used for training GPLDA). 

\subsection{Score normalization}
\label{subsec:score_norm}

For the cosine distance scoring, we used adaptive symmetric score normalization (adapt S-norm) which computes an average of normalized scores from Z-norm and T-norm~\cite{PLDA:kenny,ICSLP:Matejka}. 
In its adaptive version~\cite{ICSLP:Matejka,ICASSP2005:Sturim,Odyssey2006:Zigel}, only part of the cohort is selected to compute mean and variance for normalization. Usually $X$ top scoring or most similar files are selected;  we set $X$ to  300 for all experiments.
We created  the cohort by averaging x-vectors for each speaker in PLDA training data. It consisted of 5994 speaker models.

\subsection{Calibration and Fusion}
\label{sec:fusion}

\subsubsection{Fixed condition}

As 
we did not have any data to train the fusion on for fixed condition, we performed the fusion by computing the weighted average of the scores of four selected systems. The weights were hand-picked based on the performance of the individual systems. Also, the weights were used to compensate for the difference of the range of the scores for different backends. In particular, the highest weights of 0.4 were given to the two ResNet embedding with the cosine distance scoring systems. The other two systems had equal weights of 0.1.

\subsubsection{Open condition}

For the open condition, we trained the fusion on the VoxCeleb1\_O trials. The scores of all systems were first pre-calibrated and then passed into the fusion. The output of the fusion was then again re-calibrated. Calibration and fusion was trained by the means of logistic regression optimizing the cross-entropy between the hypothesized and true labels. The parameters optimized during the fusion were single scalar offset and the scalar combination of system weights.

\section{Results and Analysis}
\label{sec:results}

The results of the systems that went into final fusion are displayed in Table~\ref{tab:results}.
The first section of the table (lines 1-4) corresponds to the systems eligible for the fixed condition, they have seen only VoxCeleb2 dataset during the training. As our final submission for the fixed condition, we used the fusion of these four systems. The results of the fusion are shown in  line~8 of Table \ref{tab:results}. The performance of that system on the evaluation data was 1.42\:\% EER. It is interesting to note, that our previous submission was a fusion of two systems, in particular systems 1 and 3, and the performance on the evaluation set was 1.49\% EER. So, there was a marginal improvement from including two more components into the final fusion but the results did not improve dramatically. Also, one can notice that we could not gain much by training the fusion with logistic regression (system 9) instead of computing a simple weighted average (system 8).

\begin{table*}[!th]
\renewcommand{\arraystretch}{0.85}
\caption{\label{tab:results}Results of the systems on Voxceleb challenge. Cosine distance and PLDA are used as backends for ResNet and TDNN systems, respectively. Note that, for the open systems, VoxCeleb1 development data was used for training the embedding networks. That explains their good performance on E and H conditions where they are a subset of this development set.}
  \centerline{
	\setlength\tabcolsep{4pt}
    \begin{tabular}{c c c c c c   c c   c c    c c   } 
    \toprule
    \# & Fixed/Open& Acc. features& Embd NN & Backend & S-norm & \multicolumn{2}{c}{Vox1 O cleaned} & \multicolumn{2}{c}{Vox1 E cleaned} & \multicolumn{2}{c}{Vox1 H cleaned}\\
    &  & & &  & & MinDCF & EER & MinDCF & EER & MinDCF & EER\\
    \midrule \midrule
1 &Fixed& FBANK  & ResNet256 + AAM & cos       & yes        & 0.166 & 1.42 & 0.164 & 1.35 & 0.233 &2.48  \\
2 & Fixed& FBANK & ResNet160 + AAM &    cos   &    yes   & 0.154 & 1.31 & 0.163 &1.38 & 0.233 &2.50  \\
3 & Fixed& FBANK & TDNN + AAM  &   PLDA    &  no & 0.181 & 1.46 & 0.185 &  1.57 & 0.299 & 2.89 \\
4 & Fixed &PLP & TDNN     &   PLDA    & no  & 0.213 & 1.94 &  0.239& 2.03 & 0.379 & 3.97 \\
\midrule
5 & Open & FBANK & ResNet256 + AAM & cos & yes    & 0.157 & 1.22 & 0.102 & 0.81 & 0.164 &1.50  \\
6 & Open & FBANK &TDNN  & PLDA &no     & 0.195 & 1.65 & 0.170 &1.42& 0.288 &2.70   \\
7 & Open & PLP& TDNN   & PLDA &no     & 0.210 & 1.98 & 0.163 &  1.51 & 0.249 & 2.83 \\
\midrule
8 & Fixed &\multicolumn{4}{c}{ Fusion 1+2+3+4 (weighted average)  }&  0.131 & 1.02 & 0.138 & 1.14  & 0.212 & 2.12 \\
9 & Open &\multicolumn{4}{c}{Fusion 1+2+3+4 LR  } & 0.131 & 1.02 & 0.138 & 1.14  & 0.212 & 2.12\\
10 & Open & \multicolumn{4}{c}{Fusion 2+5+6+7 LR   } & 0.118 & 0.96 & 0.098 & 0.80  & 0.160 & 1.51\\
    \bottomrule  
   \end{tabular}
  }
\end{table*}

The second section of Table \ref{tab:results} shows the results of the individual systems trained for the open condition. The systems were trained using both VoxCeleb1 and 2 as well as LibriSpeech and DeepMine databases for systems 5 and 6. One should remember that when looking at the performance of these systems on  Vox1-E and Vox1-H conditions. Good results are explained by the fact that embedding extraction networks saw the test data during training. As the final submission, we used the fusion of these three systems and also included one of the fixed (ResNet160) systems. The result of the Vox1-O condition of the fusion cannot be completely reliable since we trained the fusion parameters on it. The final performance of our fusion for the open condition on the evaluation set was 1.26\:\% EER. 

Another thing to note is that our submissions for the fixed and open conditions were very similar. The main difference was in additional training data used for the open condition systems, which we believe is the reason for improved performance of the open fusion compared to the fixed one.

\section{Acknowledgements}
\label{sec:ack}
The work was supported by Czech Ministry of Interior project No. VI20152020025 "DRAPAK", Google Faculty Research Award program, Czech Science Foundation under project No. GJ17-23870Y, Czech National Science Foundation (GACR) project NEUREM3 No. 19-26934X, and by Czech Ministry of Education, Youth and Sports from the National Programme of Sustainability (NPU II) project "IT4Innovations excellence in science - LQ1602". 
\bibliographystyle{IEEEbib}

\bibliography{main}

\end{document}